\documentclass[aps,pre,twocolumn,groupedaddress]{revtex4-1}
\usepackage{amsmath,amssymb,subfigure,fancybox,txfonts,bm}
\usepackage[dvipdfmx]{graphicx}
\begin{document} 
\title{On the origin  of atomistic mechanism of rapid diffusion in alkali halide nanoclusters}

\newcommand{\vc}{\mathbf}
\newcommand{\del}[2]{\frac{\partial #1}{\partial #2}}
\newcommand{\gvc}[1]{\mbox{\boldmath $#1$}}
\newcommand{\fracd}[2]{\frac{\displaystyle #1}{\displaystyle #2}}
\newcommand{\ave}[1]{\left< #1 \right>}

\newcommand{\subti}[1]{\begin{itemize} \item {\bf #1} \end{itemize}}
\newcommand{\secti}[1]{\section{#1}}

\newcommand{\SA}{SA}
\newcommand{\SM}{SM}
\newcommand{\AH}{AH}
\newcommand{\SPM}{Surface Peeling Mechanism}
\newcommand{\VM}{Vacancy Mechanism}

\newcommand{\red}[1]{\textcolor{red}{#1}}
\newcommand{\blue}[1]{\textcolor{blue}{#1}}
\newcommand{\green}[1]{\textcolor[rgb]{0,0.6,0}{#1}}

\newcommand{\cube}[2]{#1 \times #1 \times #1 - #2}
\def\ri100{r_i^c} %
\def\KClBrxfivemthree{(KCl)$_{14}$(KBr)$_{47}$}
\def\KClBrxsevenmthree{(KCl)$_{62}$(KBr)$_{108}$}
\def\KClBrxninemthree{(KCl)$_{172}$(KBr)$_{191}$}
\def\NaClBrxfivemthree{(NaCl)$_{14}$(NaBr)$_{47}$}
\def\RbClBrxfivemthree{(RbCl)$_{14}$(RbBr)$_{47}$}
\def\KBrIxfivemthree{(KBr)$_{14}$(KI)$_{47}$}

\author{Tomoaki Niiyama}
\email{ni-yama@ike-dyn.ritsumei.ac.jp}
\author{Shin-ichi Sawada$\dagger$}
\author{Kensuke S. Ikeda$\ddagger$}
\author{Yasushi Shimizu$\ddagger$}

\affiliation{
College of Science and Engineering, 
Kanazawa University, Kakuma-machi,  Kanazawa 920-1192, Japan\\
$\dagger$ 
Department of Physics, 
Kwansei Gakuin University,Gakuen 2-1, Sanda 669-1337, Japan\\
$\ddagger$ 
Department of Physics, 
Ritsumeikan University, Noji-higashi 1-1-1, Kusatsu 525-8577, Japan
}

\begin{abstract}
To elucidate the atomistic diffusion mechanism
responsible for the rapid diffusion in alkali halide nano particles,
called {\it Spontaneous Mixing},
we execute molecular dynamics simulations
with empirical models for KCl-KBr, NaCl-NaBr, RbCl-RbBr and KBr-KI.
We successfully  reproduce essential features of
the rapid diffusion phenomenon.
It is numerically confirmed that
the rate of the diffusion clearly depends on the size and temperature
of the clusters, which is consistent with experiments. 
A quite conspicuous feature is that the surface melting
and collective motions of ions are inhibited in alkali halide clusters.
This result indicates that  {\it the Surface Peeling Mechanism},
which is responsible for the spontaneous alloying of binary metals,
does not play a dominant role for the spontaneous mixing in alkali halide
nanoclusters.
Detailed analysis of atomic motion inside the clusters reveals
that {\it the Vacancy Mechanism} is the most important mechanism
for the rapid diffusion in alkali halide clusters.
This is also confirmed by evaluation of the vacancy formation energy: 
the formation energy notably decreases with the cluster size,
which makes vacancy formation easier and diffusion more rapid
in small alkali halide clusters.
\end{abstract}

\maketitle
\section{Introduction}
\label{intro}

\label{Sec:Intro}

Over the past twenty years, a considerable amount of research of
nano size materials, e.g., carbon nanotubes, quantum dots,
nano particles, has been done 
owing  to  its technological  importance
\cite{wolf2008nanophysics,ozin2009nanochemistry}.
As is well known, nano particles play a key role for applications 
such as catalytic reactions in fuel cells 
\cite{Haruta1987AuCatalyst,Haruta:2005fk}.
Though these nano size materials are widely manipulated and synthesized, 
it has been unclear how the materials are dynamically organized microscopically.
In other words, fundamental dynamical aspects of nano materials such as
transport mechanisms of atoms in small and 
finite materials have not been fully understood yet.

Meanwhile, the {\em rapid diffusion}  which is a 
characteristic dynamical behavior peculiar to nano materials
has been found experimentally in nano metal and alkali halide particles 
\cite{yasuda1994csd,yasuda1992sos,yasuda1992saz,kimura2000AHSM,kimura2002smb}.
The experimental results reveal that 
the solid diffusion in nano clusters occurs considerably faster 
than the diffusion in bulk solids, 
and that such a rapid diffusion is 
enhanced by the smallness of the system.
Elucidation of the atomistic process 
underlying this phenomenon lead us to 
a deeper understanding 
of the atomic motion in nano size material with a surface.


The rapid diffusion in metal nano particles, which is called 
{\em Spontaneous Alloying} (SA), 
has been experimentally observed {\it in situ}, for the first time, 
using transmission electron microscopy by Yasuda and Mori 
\cite{yasuda1994csd,yasuda1992sos,yasuda1992saz}.
According to their observations,
{\SA} is the phenomenon that
nano size core-shell type bimetallic clusters prepared 
by vacuum deposition of metal atoms onto 
a cluster of another kind of metal,
e.g., gold clusters covered with copper atoms,
quickly transform into homogeneous alloy clusters 
without melting even at room temperature.
The time needed for the clusters to be alloyed
is much shorter than that of bulk bimetals.
A rough estimation tells that
the diffusion of Au in nano clusters is nine orders of magnitude 
faster than that in bulk systems.

Shimizu and his coworkers have studied this very peculiar diffusion
phenomenon by Molecular Dynamics (MD) simulations using 
two dimensional Morse models.
They suggested that surface melting peculiar to nano metal clusters
plays a crucial role in the onset of {\SA} \cite{shimizu1998cds,shimizu2001sab}.
It is well-known that the surface melting in a nano particle occurs even 
at a temperature considerably 
lower than its melting point.
In their MDs, the surface melting was observed as 
frequent rearrangements of the surface atoms 
\cite{garrigos1989mlp,PhysRevB.57.13430}.
They paid particular attention to the motion of core atoms 
which were initially encapsulated 
inside the cluster.
They found that surface melting
enables these core atoms to move collectively
without breaking into pieces 
and thus expose themselves to the surface. 
The collection of core atoms is dispersed by the surface melting 
when it becomes part of the surface.  
The mixing process of the core atoms with the surface atoms caused
by such an atomic motion is called
as {\em Surface Peeling Mechanism (SPM)}
\cite{shimizu1998cds,shimizu2001sab}.
Moreover, it has been verified that the SPM similarly 
occurs in many-body potential models different
from Morse type models.
\cite{kobayashi2002ids,kobayashi2003akt,sawada2003mds}.
Restructuring of cluster shapes by complex collective motions
has also been reported in growth simulations of binary metal clusters
with  many-body potential \cite{Baletto:2003fk}.

\begin{figure}[tbp]
\centering
\includegraphics[width=8cm]{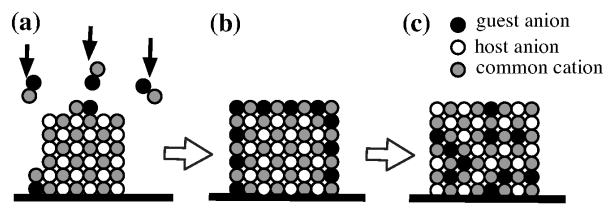}
 \caption{\footnotesize
 Schematic illustrations for the process of Spontaneous Mixing.
 (a) After an AH nano cluster consists of  host anion
 (white circles) and cations (grey circles) was prepared
 on a substrate,  another kind of AH consist of 
 guest (black circles) anion and common cations
 are deposited on the cluster by vacuum evaporation.
 (b) The AH cluster composed by host anions(white) and cations(grey) is 
 gradually covered by another kind of AH having a common cation.
 (c) These ions of the cluster rapidly mix with each other
 at room temperature when the size of initial AH cluster is sufficiently small.
 }
 \label{fig:SM-concept}
\end{figure}

After Yasuda and Mori, a similar rapid diffusion has also been 
observed experimentally in alkali halide (AH) nano clusters
by Kimura and his coworkers,
and they called it {\em Spontaneous Mixing (SM) phenomenon}
\cite{kimura2000AHSM,kimura2002smb}.
As shown by the schematic pictures of 
the phenomenon in figure \ref{fig:SM-concept},
{\SM} is rapid intermixing of AH clusters covered with
another pair of alkali halide at room temperature.
Kimura {\it et al.} found that
{\SM} occurs even in {\AH} clusters as large as about
$200$ nano meter (nm) in size.
This is much larger than the maximum size of metal clusters
for which {\SA} is observed.
They pointed out the similarities between {\SA} and {\SM}, 
since they are both considered as a manifestation of rapid diffusion 
of constituent atoms peculiar to small clusters.
However, there exists an essential difference between them.  
Kimura{\it et al} concluded that one of the most important 
controlling factors
of {\SM} is the minimum ratio of ionic radii contained in 
{\AH} clusters \cite{kimura2000AHSM,kimura2002smb}.
However, at a microscopic level, the rapid 
diffusion mechanism of constituent atoms has not been clarified yet.

In this paper, we investigate
the similarities and the essential differences between {\SM} 
in AH clusters and {\SA} in metal clusters at an atomistic level.
In particular, we are much interested in whether the {\SPM}, 
which plays the key role in {\SA}, is similarly important for {\SM}, or not.

In section \ref{Sec:Method}, our numerical model together with 
our method of research are described, and the initial configurations and 
an indicator of intermixing, which are employed in the MD simulation,
are introduced.
In Sec.~\ref{Sec:results-1}, 
we show the results for the direct observation of the MDs, and 
quantitative analyses for the intermixing process, namely, 
the dependence upon temperature, cluster size, and the
species of ions forming AH clusters are examined, paying
particular care to checking whether the cluster is melting or not.
We investigate the diffusion mechanism in Sec~\ref{Sec:results-2}
by observation of the motion of ions and vacancies inside the clusters,
then show that ions in the clusters diffused by the Vacancy Mechanism.
Finally, in Sec~\ref{Sec:discussion}, 
we  discuss the rate of the diffusion observed in our MDs,
and the reason for the rapidness of diffusion in small {\AH} clusters.

\section{Model and Numerical method}
\label{Sec:Method}

\subsection{Model}
\label{Subsec:model}

As stated in the introduction we investigate the {\SM}
by constant energy MD simulations.
For this purpose,
we consider the Hamiltonian which consists of $N$ particles
interacting with a two-body potential energy function $v(r_{ij})$
\begin{align}
H = \sum_{i=1}^{N} \frac{{\vc{p}_i}^2}{2m_i} 
+ \sum_{i<j}^N v(r_{ij}),
\label{Hamiltonian}
\end{align}
where $r_{ij}$ is the distance between $i$-th and $j$-th ions,
$|\vc{q}_i - \vc{q}_j|$,
$\vc{q}_i$ and $\vc{p}_i$ are the configurational and momentum 
coordinate of $i$-th atom, $m_i$ is the mass of $i$-th atom, respectively.
For AH clusters,
we employed the Coulomb plus Born-Mayer potential \cite{born1932gi}
given by
\begin{align}
v(r_{ij}) = {q_i q_j}/{r_{ij}} 
 + A_{k \ell} \exp{\left[({R_k+R_{\ell} -r_{ij}})/{\rho_{k \ell}}\right]},
\label{vij_CBM}
\end{align}
where $q_i$ indicates the charge of the $i$-th ion,
$k$ and $\ell$ denote species of the $i$-th and $j$-th ions,
and $A_{k \ell}$, $R_k$ and $\rho_{k \ell}$ are 
the Tosi-Fumi's parameters \cite{fumi1964isa,tosi1964isa}, respectively.
Since the original Tosi-Fumi model 
is applicable only for the single-species AH,
we extend the Tosi-Fumi potential so as to be applicable to binary AH. 
That is, we used the arithmetic mean of two different 
anion's parameters $\rho_{k \ell}$
as the parameter between these different pair.
For instance,$(\rho_{Cl-Cl} + \rho_{Br-Br})/2$ is used for $\rho_{Cl-Br}$,
where $\rho_{Cl-Cl}$ and $\rho_{Br-Br}$ are the parameters for 
KCl and KBr, respectively.
(Note that the value of the parameters, $\rho_{k \ell}$, 
for cation-cation and anion-anion interaction are 
assumed to be equal to the value of the parameter 
for cation-anion interaction in the Tosi-Fumi's model 
\cite{fumi1964isa,tosi1964isa}.)
In this paper, we denote an {\AH} cluster consisting of 
alkali halides AB and CD  as AB-CD, or 
(AB)$_n$(CD)$_m$ when $n$ AB and $m$ CD pairs
constitute the cluster: for instance, 
NaCl-KBr or (NaCl)$_n$(KBr)$_m$.

For simplicity, we consider AH clusters which consist of  
only two different kind of anions and common cations 
e.g., KCl clusters covered with KBr ions.
Now we name the anions inside the cluster like Cl as {\em host}, while we call 
the anions which are initially located at the surface like Br as {\em guest}.
And we call such a cluster a {\em binary cluster},
because we are concerned with the intermixing between host and guest anions.
In Fig.~\ref{fig:SM-concept}, cations, host anions and guest anions are
shown by grey, white and black circles, respectively.

\subsection{Initial configurations and numerical method}
\label{Subsec:init-conf}

\begin{figure}[tbp]
\centering
\includegraphics[width=8cm]{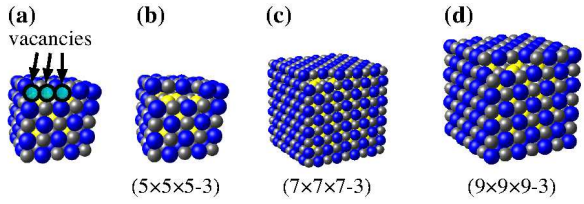} 
\caption{\footnotesize 
Typical initial configurations for the MD simulations of AH clusters.
Three ions were removed from one of the edges of 
$(n \times n \times n)$ shape AH clusters, i.e.,
there are three atomic vacancies on the edge,
where guest, host anions and common cations are colored 
blue, yellow and grey.
}
\label{fig:init-conf}
\end{figure}

Next we explain the configurations of
ions in a cluster at the initial time ($t=0$).
We employed the configurations
in which several ions were removed from the cubic cluster
with simple cubic lattice, specifically
$(5 \times 5 \times 5 - 3)$,
$(7 \times 7 \times 7 - 3)$ and 
$(9 \times 9 \times 9 - 3)$ structure (see Fig \ref{fig:init-conf}),
where $(n \times n \times n - \alpha)$ denotes the resultant 
structure where $\alpha$ ions were removed from 
an edge of a cluster forming cubic
structure composed of $n \times n \times n$ atoms.
The reason for such a choice of initial configuration
is that atomic vacancies and their migration are indispensable
for intermixing in AH clusters to be initiated, as will be
described later.

It is a difficult problem to prepare an ``ideal'' structure of
the initial configuration for rapid diffusion, because
the natural configurations of atoms in the initial stage 
has not been clarified in any experiments.
If one chooses a {\it close-packed structure} as the initial
configuration, [the $(n \times n \times n)$ cubic structure
is one of the close-packed structures of AH nano particles which is
quite stable], it is easily expected that no intermixing
phenomenon will be observed.
Due to the stability of the closed-packed structure
one has to wait a very long time for any atom to migrate 
even to its nearest neighbor site.
For this reason, we employed non-close-packed structures.
(The stability and reliability of this type of non-close-packed structure
has been discussed in experiments and ab initio calculations 
\cite{Aguado2001AbInitioClusters}.)

To constitute an initial configuration, we firstly took 
a $[(n-2) \times (n-2) \times (n-2)]$ cubic cluster
which is constituted of host anions and cations. Next, 
we cover the cluster with a mono layer of alkali halides
which consist of guest anions and cations.
Finally some ions ( the total number is $\alpha$ ) were removed from 
one of the twelve edges of the cubic cluster.
We chose $\alpha = 3$ in this study.
As a result, we prepared an initial configuration 
with three neighboring vacancies on a surface edge,
as shown in Fig.~\ref{fig:init-conf}.

\label{Subsec:method}

We executed MDs under the isoenergetic condition
by the velocity form of the Verlet algorithm,
where the time step was typically $10$ femto seconds.
Numerical results described below are mainly based 
on the simulation of KCl-KBr clusters.
We also mention some results of the simulation for 
NaCl-NaBr, RbCl-RbBr and KBr-KI clusters.
Note that Cl and Br are respectively
host and guest anions in KCl-KBr, NaCl-NaBr and RbCl-RbBr experiments, 
and that Br and I  are respectively host and guest anions in KBr-KI experiments.

First of all, employing the velocity scaling method in time evolution
we raised the temperature of a cluster
having the initial configuration shown in Fig.~\ref{fig:init-conf}
from ground state to a desired temperature.
Here, we used the micro canonical kinetic temperature $T$ defined by
\begin{align}
T = \frac{2 \ave{K}}{(3N-6) k_B},
\label{kinetic-temperature-def}
\end{align}
where $\ave{K}$, $N$ and $k_B$ are time-average of the total kinetic
energy, the total number of the atoms and 
Boltzmann's constant, respectively \cite{jellinek1986slp,allen1989csl}.
The factor $6$ in the above equation results from
the conservation of 
the total translational and angular momentum of the system
under the isolated condition \cite{niiyama:051101}.

Then, after achieving the desired temperature, the cluster was
evolved under the isoenergetic condition,
where the total simulation time periods were from $200$ 
to $2000$ nano seconds (ns) depending on the situation.
In each MD run, temperature $T$ indicates an initial value 
of the micro canonical kinetic temperature.
Since vibrational period is typically $10^{-1} - 10^0$ pico seconds (ps),
we averaged every time series of the cluster configuration
over a period of $10$ ps, finally obtaining a series of configuration
from which thermal vibration of the atoms was almost eliminated.

\subsection{Indicator of mixing}
\label{Subsec:invasion-deg}
\newcommand{\inv}{R_g}
\newcommand{\invf}[1]{R_g(#1)}

In order to quantify the time evolution
of the mixing process, let us 
introduce an indicator of intermixing.
The intermixing requires motion of the guest atoms toward
the center of a cluster, so we can use a measure of the average
distance of the guest atoms from the center of the cluster
as an indicator of mixing.
If a cluster has spherical symmetry, the average distance of 
guest atoms from the center of a cluster, 
$r_g = \frac{1}{N_g} \sum_{i=1}^{N_g} r_i$,
where $r_i = | \vc{q}_i - \vc{R}|$
and $\vc{R}$ is the position of the center of the cluster, is
suitable for an indicator of intermixing
\cite{shimizu1998cds,shimizu2001sab,kobayashi2002ids}. 
However, the present clusters have cubic symmetry. 
The value of $r_g$ may vary even by a trivial movement of atoms
along the surface such as surface diffusion,
which can not be regarded as the intermixing.
Therefore, we introduce another definition of distance
that is the maximum of the distances from the center along
the axes, defined by
\begin{align}
\ri100  = |\vc{q}_i - \vc{R}| \equiv 
\max \left\{ |(\vc{q}_i - \vc{R})\cdot \vc{e}_j|, j = 1, 2, 3 \right\},
\label{ri100-definition-sawada}
\end{align}
where $\vc{e}_1$, $\vc{e}_2$ and $\vc{e}_3$
are the three independent vectors
which are along the crystal axis and
perpendicular to three surfaces of a cubic cluster.
We shall call this the ``principal radial distance''.

\begin{figure}[tbp]
\centering
\includegraphics[width=8cm]{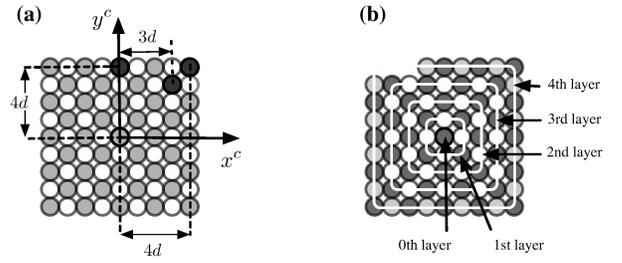}
 \caption{\footnotesize
 A schematic figure of an AH cluster is represented 
 by a two-dimensional square cluster.
 (a) A guest anion is colored by black, 
 and host anions and common cations are represented 
 by grey and white circles, respectively.
 In the case that the guest anion which is initially located 
 at $(x^c,y^c)=(4d,4d)$ occasionally hops along the surface
 to the lattice site $(x_c,y_c)=(0,4d)$ (surface diffusion),
 the distance, $\ri100$, keeps the constant value 
 $4d$  during the hopping process.
 However, the distance, $\ri100$ decreases to $3d$, when 
 the initial guest anion hops to an inside site, e.g., $(x_c,y_c)=(3d,3d)$.
 (b) 
 We intentionally introduce atomic vacancies 
 on the cluster surface by removing some atoms.  
 The white lines indicate the layers which virtually decompose 
 a cluster into parts. 
 We define the index for the layers, which is allocated 
 by ascending order from the surface, 
 because it is helpful for a simple indication for the penetration depth of 
 the guest anions (See SubSec.~\ref{Subsec:T-depend}).
}
 \label{fig:invade100-model}
\end{figure}

In Fig.~\ref{fig:invade100-model}~(b),
a typical two dimensional example is shown.
Each layer is denoted by  $0$th, $1$st,
$\cdots$, $\ell$-th, $\cdots$, $L$-th layer 
from the center to the surface. The integer $\ell$ is the index of a layer.
Note that $0$th and $L$-th layer correspond to the center 
and surface of the cluster, respectively.
When the $i$-th guest anion in a cubic cluster remains 
in the vicinity of its lattice point in the $\ell$-th layer, 
the value of $\ri100$ is $\ell d$.
Using the layer index, the average distance $r_g$ is represented by
$r_g = \frac{1}{N_g} \sum_{\ell=0}^{L} n_{\ell} \ell d$,
where $n_{\ell}$ and $N_g$ is the number of the guest anions
in the $\ell$-th layer and the number of all guest atoms, respectively.
Note that the value $r_g$ does not vary whenever any guest atoms
move along a layer [see also Fig.~\ref{fig:invade100-model}~(a)].
Normalizing the average distance $r_g$ by $Ld$,
we here define the value $\inv$ as an indicator of intermixing:
\begin{align}
\invf{t} \equiv \frac{r_g}{L d} = 
\frac{1}{N_g L} \sum_{\ell=0}^{L} n_{\ell}(t) \ell,
\label{discrete-inv-deg}
\end{align}
%
where $n_{\ell}(t)$ is the number of the guest atoms belonging to
$\ell$-th layer at time $t$.

In order to quantify the process of mixing, we estimate 
the ideal value for $\inv$ that will be realized
if the host and guest anions are homogeneously mixed.
For this, let us assume
a single AH cluster has $(N_g+N_h)$ lattice sites,
where  $N_g$ and $N_h$ are numbers of guest and host anions.
We also assume that the guest anion occupies one of $(N_g+N_h)$
lattice sites with equal probability.
Consequently the probability for finding a guest atom 
on an arbitrary lattice site for anions is 
$N_g/(N_g+N_h)$ when guest anions are spread homogeneously
throughout a cluster by stochastic jumps.
Thus, in such a mixed cluster the average number of the guest
atoms in $\ell$-th layer is represented by
$\ave{n_{\ell}}=N_{\ell} N_g/(N_g+N_h)$,
where $N_{\ell}$ is the number of the lattice sites for guests in
$\ell$-th layer.
By substituting $\ave{n_{\ell}}$ to Eq.~(\ref{discrete-inv-deg}),
one can derive the value $\inv$ in an ideal mixed cluster,
\begin{align} 
\ave{\inv} = \frac{1}{(N_g+N_h) L} \sum_{\ell=0}^L N_{\ell} \ell.
\label{ideal-inv-deg}
\end{align}

For the $(\cube{5}{3})$ cluster in Fig.~\ref{fig:init-conf},
the number of the lattice sites for guests in each layer
is $(N_0, N_1, N_2) = (0, 14, 47)$,
and $L = 2$, $N_g = 47$, $N_h = 14$,
so we obtain the ideal value $\ave{\inv} = 0.885$.
Similarly, the ideal value is $\ave{\inv} = 0.852$ 
in a $(\cube{7}{3})$ cluster, where
$(N_0, N_1, N_2, N_3) = (0, 14, 48, 108)$,
$L=3$, $N_g = 108$ and $N_h = 62$.
In a $(\cube{9}{3})$ cluster, 
the ideal value is $\ave{\inv} = 0.829$,
where 
$(N_0, N_1, N_2, N_3, N_4) = (0, 14, 48, 110, 189)$,
$L=4$, $N_g = 191$ and $N_h = 172$.
By computing the value of the indicator $\inv$ 
[Eq.~(\ref{discrete-inv-deg})] with the value $\ave\inv$
for an ideal mixed cluster [Eq.~(\ref{ideal-inv-deg})],
one can quantify the
degree of mixing during the SM.


\secti{Direct observation and qualitative analysis}
\label{Sec:results-1}

\subsection{Direct observation of intermixing in clusters}
\label{Subsec:direct-observation}

We show that AH clusters spontaneously mix
without melting in the present MD simulations, 
where we employed KCl-KBr clusters:
\KClBrxfivemthree, \KClBrxsevenmthree and \KClBrxninemthree 
whose initial configurations are 
($\cube{5}{3}$), ($\cube{7}{3}$) and ($\cube{9}{3}$) structure
as described in Subsec.~\ref{Subsec:init-conf}, respectively.
We chose the present AH cluster, because the {\SM} 
for KCl-KBr clusters has been observed
in the experiments \cite{kimura2000AHSM,kimura2002smb}.
We executed isoenergetic MDs for \KClBrxfivemthree at $650$ K 
($E = -3.0985 \pm 0.0002$ eV/atom),
\KClBrxsevenmthree at $700$ K ($E = -3.1441 \pm 0.0002$ eV/atom)
and 
\KClBrxninemthree at $800$ K ($E = -3.1468 \pm 0.0002$ eV/atom),
where $E$ is the total energy of the system.

\begin{figure}[tbp]
\centering
\includegraphics[width=8cm]{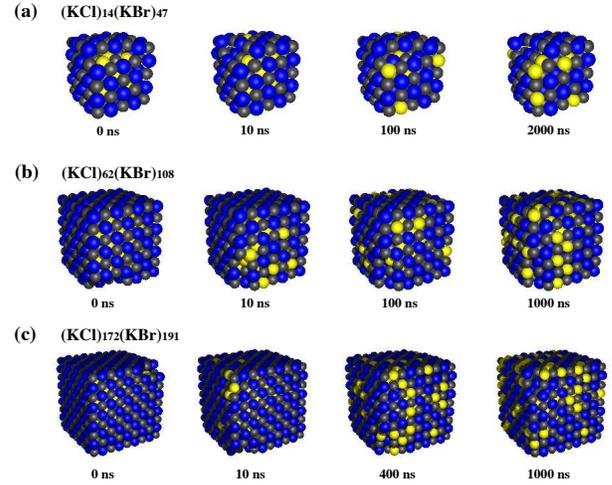}
 \caption{\footnotesize
 Snapshots of the configuration of 
 \KClBrxfivemthree, \KClBrxsevenmthree and \KClBrxninemthree cluster
 in time evolution at $650, 700$ and $800$ K, respectively.
 The yellow-colored host anions (Cl$^-$), which were located inside
 the clusters at $t=0$ gradually appeared on the surface in these
 clusters.
 This is a manifestation of the fact that host anions 
exchange positions with guest anions
(Br$^-$) which are colored by blue.
 }
 \label{fig:KClBr-snapshot-all}
\end{figure}

The snapshots taken at $10$ ps intervals
are shown in Fig.~\ref{fig:KClBr-snapshot-all}.
As is evident from the figure, host anions (Cl$^-$, colored yellow) 
which have been located on the cluster's inside at $t=0$
gradually aggregate on the surface by
exchanging positions with guest anions (Br$^-$, colored blue)
located on the surface at $t=0$.
During the entire time evolutions the clusters keep a simple cubic shape
\footnote{
Note that \KClBrxfivemthree rarely causes large-scale structural transition
from a simple cubic shape to a rectangular parallelepiped 
by shear deformation along a diagonal line on a surface.
But such events are so rare that we ignored several MD runs including 
these events
(in fact, we cannot observe the events in larger clusters,
\KClBrxsevenmthree and \KClBrxninemthree).
}.
On the other hand, the three vacancies that are initially neighbors
on a surface edge at $t=0$ tend to move together on the edges of 
the cluster, with occasional independent excursions inside of 
the cluster.
(One can see the vacancies begin to move separately 
in Fig.~\ref{fig:KClBr-snapshot-all}~(b) at $t=10$ ns).


%
%
\begin{figure}[tbp]
\centering
\includegraphics[width=8cm]{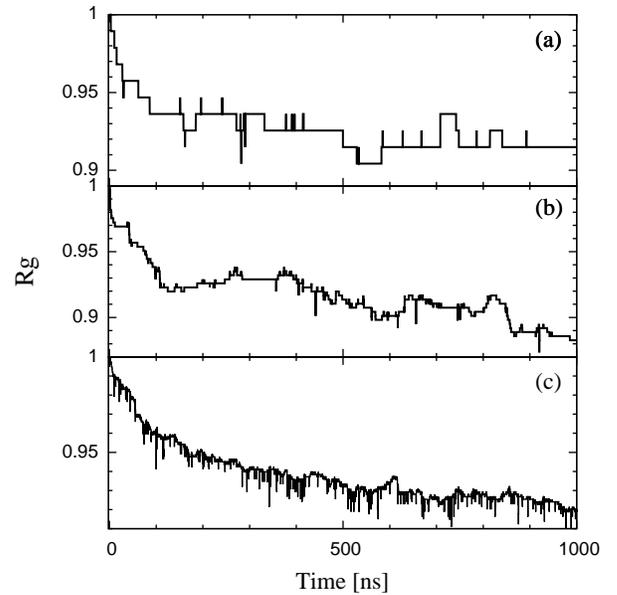}
 \caption{\footnotesize 
 Typical time evolutions of $R_g(t)$ for three different sizes
of clusters. 
The initial temperature is higher for larger cluster size.
(OR) The larger the cluster size, the higher the initial temperature.
 (a) \KClBrxfivemthree at $650$ K.
 (b) \KClBrxsevenmthree at $700$ K.
 (c) \KClBrxninemthree at $800$ K.
 }
 \label{fig:inv-deg-oneSamples}
\end{figure}

Next we provide quantitative description of the progress of {\SM} 
in the KCl-KBr clusters using the indicator $\inv$ 
[Eq.~(\ref{discrete-inv-deg})].
In Fig.~\ref{fig:inv-deg-oneSamples}~(a), we show typical time
evolutions of $\inv(t)$ of \KClBrxfivemthree at $T=650$ K, \KClBrxsevenmthree at $T=700$ K
and \KClBrxninemthree at $T=800$ K, corresponding to the case in
Fig.~\ref{fig:KClBr-snapshot-all}.
In Fig.~\ref{fig:inv-deg-oneSamples}
one can see that the value $R_g$ is unity at $t=0$.
The initial value is $\inv(0) = 1$ 
by the definition.
The value $R_g$ decreases discontinuously with small spiky fluctuations 
as is seen in Fig.~\ref{fig:inv-deg-oneSamples}.
This fluctuation can be smeared out
by averaging with several samples of MD runs
as shown in Fig.~\ref{fig:ave-inv-T-depend-KClBr}.
The resultant averaged value shows a monotonous decreasing trend.
However, the indicator $R_g(t)$ did not reach its ideal value
$\ave{\inv} = 0.885$ over the time evolution.
The behaviors of the other size clusters shown in 
Fig.~\ref{fig:inv-deg-oneSamples}~(b) and (c) are similar to
those of \KClBrxfivemthree described in the above.
That is, the way of intermixing is common to these three types 
of AH clusters at a qualitative level. 
It is evident that the intermixing tends to be suppressed in larger clusters. 
Because the intermixing is enhanced in smaller clusters,
we  hereafter call the present intermixing Spontaneous Mixing (SM).

Such intermixing in {\AH} clusters is observed quite commonly
in simulations where host anions are initially covered with
guest anions.
It is plausible to expect that these examples are 
manifestations of the {\SM} which was experimentally observed.

\subsection{Intermixing without melting}
\label{Subsec:Lindemann}

According to the experiments for {\SA},
the completely intermixed state is attained without melting
\cite{yasuda1994csd,yasuda1992sos,yasuda1992saz}.
The completely intermixed state in {\AH} clusters 
is also attained in a solid phase \cite{kimura2000AHSM,kimura2002smb}.
It is necessary to confirm 
that the AH cluster in the present simulation is solid-like
from beginning to end during {\SM}.
The snapshots in Fig.~\ref{fig:KClBr-snapshot-all} show that
the KCl-KBr clusters keep simple cubic shapes during the intermixing.
Although it is obvious that the clusters did not melt 
by the direct observation of these snapshots, 
we give quantitative evidence of non-melting.
The presence of melting is probed by the so-called Lindemann's index,
which is a measure of the relative fluctuation of the distance 
between any two atoms as follows,
\begin{align}
\delta = \frac{2}{N(N-1)} \sum_{i<j}^N 
\frac{\sqrt{\ave{{r_{ij}}^2} - \ave{r_{ij}}^2}}{\ave{r_{ij}}},
\label{lindemann}
\end{align}
where $N$ is the number of atoms in the system, and
$\ave{r_{ij}}$ is the long-time average distance between 
$i$-th and $j$-th atom \cite{sugano1987m}.
When $\delta$ is larger than $0.1$,
the system is empirically considered to be melted.
We verify that an averaging interval of $100$ ps
is long enough for the value of the index to be well averaged.

%
\begin{figure}[tbp]
\centering
\includegraphics[width=8cm]{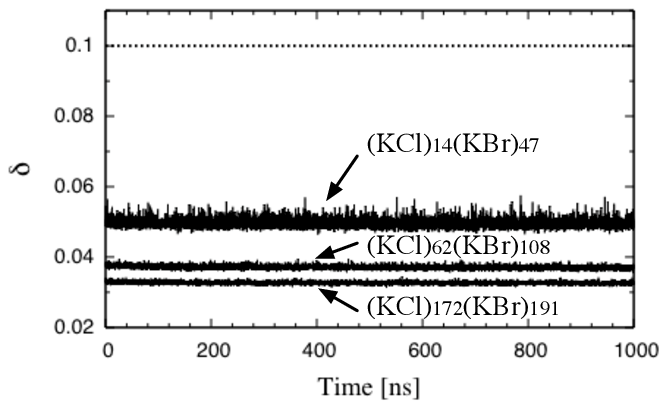}
 \caption{\footnotesize 
 Typical time evolution of Lindemann's index $\delta$ 
 during SM for \KClBrxfivemthree, \KClBrxsevenmthree and \KClBrxninemthree. 
 }
 \label{fig:Lindemann-KClBr-all}
\end{figure}

In Fig.~\ref{fig:Lindemann-KClBr-all},
we show the typical time evolution of $\delta$ during {\SM} for
\KClBrxfivemthree, \KClBrxsevenmthree and \KClBrxninemthree,
corresponding to Fig.~4 and Fig.~5.
The values of $\delta$ for 
\KClBrxfivemthree, \KClBrxsevenmthree and \KClBrxninemthree are
about $0.05$, $0.04$ and $0.03$, respectively.
These are sufficiently $smaller$ than the melting criterion $0.1$,
thus the {\SM}  for KCl-KBr clusters is considered to
proceed in solid phase.

\subsection{Dependence upon ion species}
\label{Subsec:Ion-depend}

To see the dependence of {\SM} upon the choice of ion species,
we examined several combinations of alkali halides
such as NaCl-NaBr, RbCl-RbBr and KBr-KI, in addition to KCl-KBr.
For comparison,
\NaClBrxfivemthree, \RbClBrxfivemthree and \KBrIxfivemthree with initial structure
($\cube{5}{3}$) were employed,
and the MD simulations for each cluster were done by setting the initial 
temperatures as $T=600$ K.

%
\begin{figure}[tbp]
\centering
 \includegraphics[width=8cm]{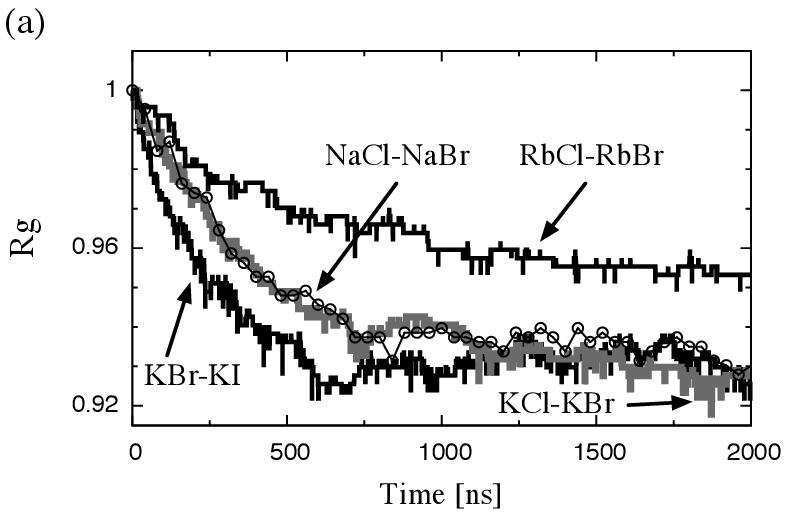}
 \includegraphics[width=8cm]{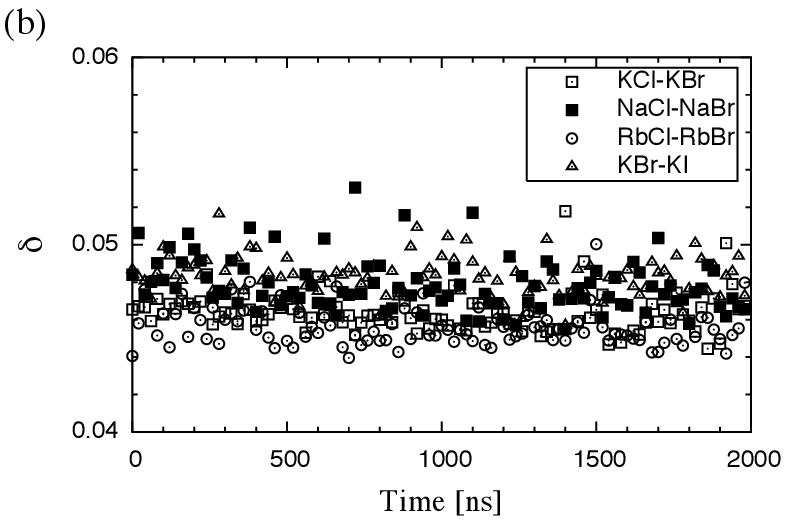}
 \caption{
 Typical time evolution of 
 (a) $\inv$ and (b) Lindemann's indices $\delta$
 for \KClBrxfivemthree, \NaClBrxfivemthree, \RbClBrxfivemthree and \KBrIxfivemthree
 at $T = 600$ K.
 }
 \label{fig:Ions-comp}
\end{figure}

The time evolution of $\inv(t)$ obtained by averaging over 5 MD runs
is depicted in Fig.~\ref{fig:Ions-comp}~(a).
The figure shows that all the values of the averaged distance $\inv$s 
decrease monotonically and their behaviors are similar to that of KCl-KBr ones.
In fact, one can verify that 
$\inv$ for \KClBrxfivemthree, \NaClBrxfivemthree and \KBrIxfivemthree  exhibits
a common decreasing rate, while 
\RbClBrxfivemthree shows a rather slow decreasing trend.
Note that $\inv$ does not reach the ideal value,
$\ave{\inv} = 0.885$, in all cases.

Typical time evolutions of Lindemann's index,
$\delta$, are depicted in Fig.~\ref{fig:Ions-comp}~(b),
where the points are plotted with $200$ ps interval.
During {\SM}, all the values of $\delta $ are about $0.05$ 
which is smaller than the melting criterion $0.1$.
Therefore, these clusters similarly show full intermixing 
in solid phase without exception.

According to the experiment for AH clusters by Kimura {\it et al.}, 
one of the dominantly controlling
factors of {\SM} is the minimum ratio 
of ionic radii of alkali halide constituent of clusters.
More precisely, the intermixing occurred in the clusters whose minimum ratio of
the ionic radii is larger than 68 \%
\cite{kimura2000AHSM,kimura2002smb}.
According to this criterion, KCl-KBr and RbCl-RbBr clusters
should exhibit intermixing,
because the ratios of radii between cation and anion 
are (K/Br = $0.68$), (Rb/Br = $0.76$) and (K/Cl = $0.73$), respectively.
On the contrary, 
one should expect the absence of intermixing
in NaCl-NaBr and KBr-KI clusters (Na/Br = $0.5$ and K/I = $0.62$), 
where the ratio of the radii of anions, Cl/Br and Br/I are $0.93$ and $0.90$,
respectively.
However, the present simulation results predict that there exists
intermixing in NaCl-NaBr cluster like in other AH clusters. 
This discrepancy suggests that
the dependence will be reversed in smaller AH clusters
such as in our MD's results,
because the clusters employed in the experiments
are much larger than those used in our simulations
(the size of clusters in the experiments was several hundred nano meters).

\subsection{Temperature-dependence}
\label{Subsec:T-depend}

According to Yasuda and Mori, {\SA} is enhanced for metal clusters,
as the temperature is raised.
However, no experimental observation showing
the role of temperature is known for {\SM} in AH clusters.
In this subsection, we probe the temperature-dependence
of intermixing in AH clusters.
We calculated the time for $\inv(t)$ 
to reach a certain value for various initial temperatures.

%
\begin{figure}[tbp]
\centering
\includegraphics[width=8cm]{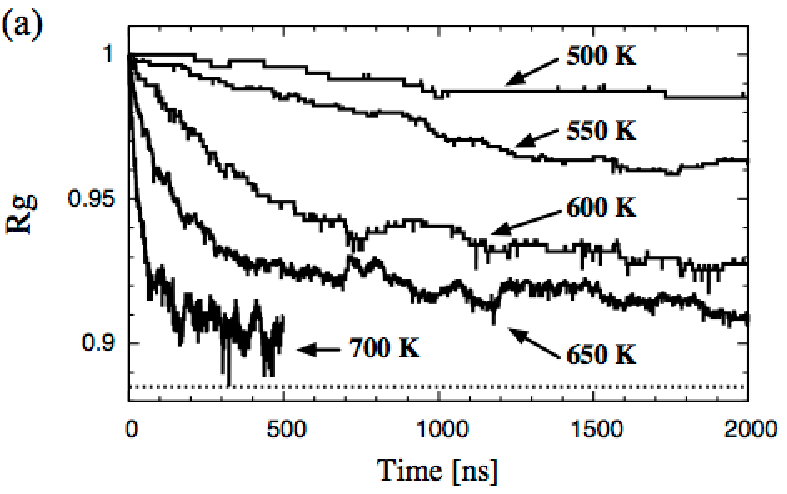}
 \includegraphics[width=8cm]{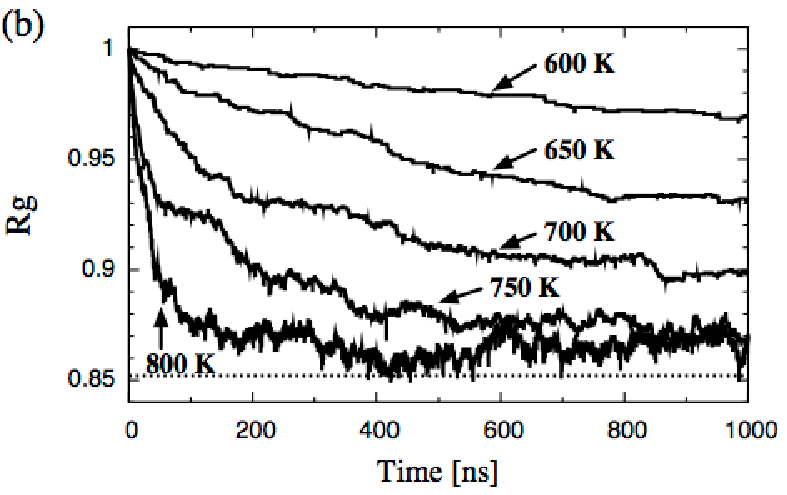}
 \includegraphics[width=8cm]{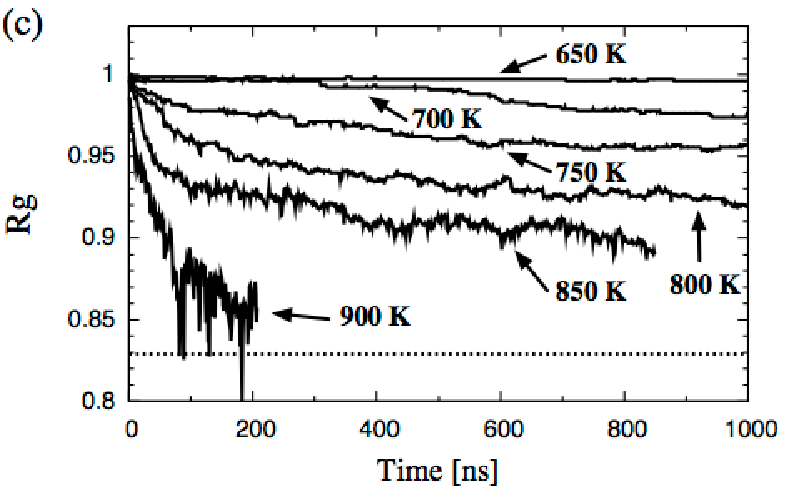}
 \caption{\footnotesize
 Time evolution of the averaged distance $R_g(t)$
 depending on various initial temperatures.
 The dot lines represent the ideal averaged distance $\ave{R_g}$ 
 corresponding to a realization of perfect intermixing 
 between host and guest anions.
 Three kinds of AH clusters are shown in 
 (a) \KClBrxfivemthree. (b) \KClBrxsevenmthree. (c) \KClBrxninemthree, respectively.
 }
 \label{fig:ave-inv-T-depend-KClBr}
\end{figure}
%
The MD trajectories for \KClBrxfivemthree at
the initial temperatures, $T=$ $500, 550, 600, 650$ and $700$ K,
were computed for five different initial conditions, respectively.
The behaviors of  $\inv(t)$ averaged over 5 samples for each temperature
are shown in Fig.~\ref{fig:ave-inv-T-depend-KClBr}~(a).
The results for
\KClBrxsevenmthree ($T=600$, $650$, $700$, $750$, $800$ K)
and \KClBrxninemthree ($T=650$, $700$, $750$, $800$, $850$, $900$ K)
are also depicted in Figs.~\ref{fig:ave-inv-T-depend-KClBr}~(b) and (c),
respectively.
The figures clearly demonstrate that $\inv(t)$ of all the clusters
tend to decrease faster as the temperature is raised.
In short, raising temperature promotes the diffusion in
binary AH clusters.
This dependence on temperature is quantitatively consistent with
the conventional  diffusion process and other studies of
metal nano particles 
\cite{shimizu1998cds,shimizu2001sab,
kobayashi2002ids,kobayashi2003akt,sawada2003mds}.
Arrhenius law for a diffusion of atoms in solid yields a 
well-known formula for the  diffusion rate $D$ as a function of 
temperature $T$: 
\begin{align}
D = D_0 \exp \left( - {E_{act}}/{k_B T} \right),
 \label{diffusion-const-eq}
\end{align}
where $D_0$, and $E_{act}$ are a frequency factor 
and the activation energy of the diffusion, 
respectively \cite{shewmon:gds,Kittel2005ed8th}.

\begin{figure}[tbp]
\centering
\includegraphics[width=8cm]{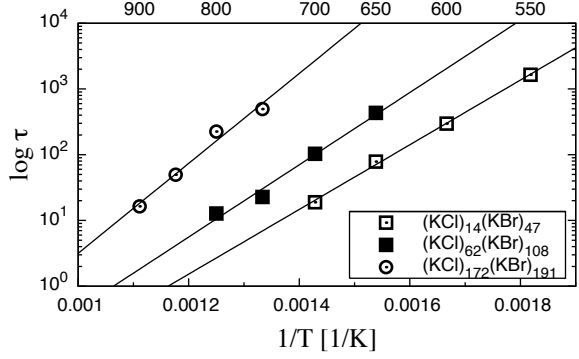}
 \caption{\footnotesize
 The temperature dependence of the mixing time $\tau$ of 
 \KClBrxfivemthree, \KClBrxsevenmthree and \KClBrxninemthree.
 The three solid lines, which represent
 Arrhenius type temperature dependence, are
 fitted by the least-squares method [Eq.~(\ref{Arrhenius-tau})].  
 }
 \label{fig:Arrhenius-plot}
\end{figure}

\def\Rncri{\inv'}

It is difficult to evaluate a diffusion rate of guest anions in clusters,
because a large portion of surface atoms 
leads to the spatial-dependence of the diffusion rate.
Instead of evaluating the diffusion rate, we focus our attention
to the time $\tau$ required for $\inv$ to reach a specific value $\Rncri$,
\begin{align}
\inv(\tau) = \Rncri.
\label{eq:Rg'}
\end{align}
We consider $\tau$ as the mixing time, and examine its temperature-dependence.
Due to the difficulty of the choice $\Rncri$,
we tentatively employed the criteria
$\Rncri = \ave{\inv} + 2/3(\inv(0)-\ave{\inv})$;
the criteria of 
\KClBrxfivemthree, \KClBrxsevenmthree and \KClBrxninemthree are
$\Rncri = 0.962$, $0.951$ and $0.943$, respectively.

The semi-log plot of the inverse temperature versus 
the mixing time $\tau$ 
is depicted in Fig.~\ref{fig:Arrhenius-plot}.
The three lines in the figure are obtained 
by assuming Arrhenius type temperature-dependence,
\begin{align}
\tau = \tau_0 \exp(E_{act}/k_B T).
 \label{Arrhenius-tau}
\end{align}
Two parameters, $E_{act}$ and the frequency factor $\tau_0$
are decided by the least-squares method.
For \KClBrxfivemthree, \KClBrxsevenmthree and \KClBrxninemthree ,
we obtained $E_{act} = 0.98$, $1.1$ and $1.4$ eV and 
$\tau_0 = 1.9 \times 10^{-6}, 1.4 \times 10^{-6}, 4.6 \times 10^{-7}$ ns,
respectively.

One can see that the behavior of  $\tau$ for the KCl-KBr clusters 
in Fig.~\ref{fig:Arrhenius-plot} well follows the Arrhenius type law by
choosing suitable values for the fitting parameters.
The present result is consistent with the previous MD studies for {\SA}
\cite{shimizu1998cds,shimizu2001sab,
kobayashi2002ids,kobayashi2003akt,sawada2003mds}.
That is, the time to accomplish the {\SM} becomes longer in an exponential way,
as the temperature is decreased, as in the case of {\SA}.


\subsection{Size-dependence}
\label{Subsec:size-depend}

As is shown in the previous section, 
{\SA} and {\SM} have similarities in their time-dependent features.
One may thus expect that other features of {\SM} are 
also similar to those of {\SA}.
However, systematic studies for {\SM} in AH cluster have been
left untouched,
whereas extensive experimental and theoretical studies 
have been done for {\SA}. 
For instance, the size-dependent feature of {\SM} has not been 
fully understood yet,
while it is well-known that {\SA} is suppressed 
for the larger metal nanoclusters.
In spite of the lack of systematic experimental studies for {\SM}, 
it is reasonable to expect that {\SM} will be suppressed 
for  larger AH clusters.

To confirm such an expectation, we examine the size-dependent property
of intermixing in KCl-KBr clusters.
In particular, we put our emphasis upon the time, $\tau_g$,
required for guest atoms to migrate over a certain fixed distance, 
say $\Delta r_g$, and examine the size-dependence of $\tau_g$.
Here we employed $0.076 d \ \AA$ as the value of $\Delta r_g$,
then evaluated $\tau_g$ for this value
from data in Fig.~\ref{fig:ave-inv-T-depend-KClBr}.

To do that, we have to remark on the relation between $\Delta r_g$ and $\inv$
because Fig.~\ref{fig:ave-inv-T-depend-KClBr} shows only the time evolutions
of $\inv$, not $\Delta r_g$.
Using average distance $r_g$ between guest atoms 
and the center of a cluster,
$\Delta r_g$ is expressed as
$\Delta r_g = r_g(0)-r_g^*$, where 
$r_g(0)$ is the distance at the initial time and 
$r_g^*$ is that at the given time.
Note that $r_g(0)$ and $r_g^*$ depend on the cluster size.
As mentioned in Subsec. \ref{Subsec:invasion-deg},
the averaged distance is represented by $r_g(t) = L d \inv(t)$,
where $d$ is the lattice constant in a simple cubic structure.
Thus, the migration length of the guest anions is alternatively expressed
in terms of the normalized distance $\inv(t)$ as 
\begin{align}
\Delta r_g =L d (\inv(0)-\inv^*)= L d (1 - \inv^*).
\end{align}
Let us recall that
\KClBrxfivemthree, \KClBrxsevenmthree and \KClBrxninemthree clusters
form $(\cube{5}{3})$, $(\cube{7}{3})$ and $(\cube{9}{3})$ 
structures, and the number of their layers is $L = 2,\ 3$ and $4$,
so the corresponding values of $\inv^*$ for $\Delta r_g = 0.076 d \ \AA$
are $\inv^* = 0.962,\ 0.975$ and $0.981$, respectively.
Therefore, $\tau_g$ can be translated as the time when $\inv(t)$ reaches
$0.962,\ 0.975$ and $0.981$ 
for \KClBrxfivemthree,\KClBrxsevenmthree and \KClBrxninemthree, respectively.

For evaluating $\tau_g$, we chose the MD results at $700$ K 
shown in Fig.~\ref{fig:ave-inv-T-depend-KClBr}.
With these data we obtained 
$\tau_g = 18.9$,\ 
$42.1$,\ 
and $668$ ns 
for \KClBrxfivemthree, \KClBrxsevenmthree and \KClBrxninemthree clusters, respectively.
Thus, the time required by guest anions to diffuse 
over the distance increases very rapidly with an increase 
in the size of the cluster, which means that the {\SM}
is inhibited in larger clusters.
In this sense, we verified the similarities 
in the size-dependence between {\SM} and {\SA}.

Furthermore, it is worth to emphasize that 
we also confirm the same aspect of 
the size-dependence in Fig.~\ref{fig:Arrhenius-plot}. 
As mention in the previous subsection, 
the activation energy introduced in Eq.~(\ref{Arrhenius-tau})
becomes larger for the larger AH cluster. 
In both cases, we ascertain that the size of AH cluster affects 
the mobility of guest anions
entering the center of the cluster.


\secti{Diffusion mechanism}
\label{Sec:results-2}

We showed that intermixing in AH clusters
occurred spontaneously in solid phase.
In this section, we explore some details of
the atomistic diffusion mechanism of the intermixing. 
Firstly, we attempt to quantify the onset of 
surface melting which is considered to be a necessary condition 
for the {\SPM}.
Then we characterize the motion of atoms inside a cluster 
by focusing attention on the core of the cluster.
The results of these observations imply that the diffusion mechanism 
of intermixing is the so-called {\VM}  \cite{shewmon:gds}.
Finally, by visualizing trajectories of all ions of an {\AH} cluster
we present direct evidence indicating that the diffusion 
is governed by the {\VM}.

\subsection{Surface melting}
\label{Subsec:surface-melting}

Surface melting of metal clusters
which occurs even at temperatures considerably lower than 
their melting points has been
reported in some experimental studies
\cite{Kofman1989630,PhysRevB.57.13430,garrigos1989mlp}.
As mentioned in Sec.~\ref{Sec:Intro}, the surface melting 
is one of the necessary conditions for the {\SPM} to work.
Indeed, the previous results based on 2D Morse model show how 
surface melting enhances {\SA} 
by the {\SPM} \cite{shimizu1998cds,shimizu2001sab}.
Therefore, it is plausible to expect that the same mechanism similarly works 
in  {\SM} of AH clusters.
In this subsection, we verify the absence
of surface melting on AH clusters during {\SM}.

Following the previous study \cite{shimizu2001sab},
we defined the Lindemann's index for the $i$-th atom,
\begin{align}
\delta_i = \frac{1}{N-1} \sum_{j \ne i}^N 
 \frac{\sqrt{\ave{{r_{ij}}^2}-\ave{r_{ij}}^2}}{\ave{r_{ij}}},
\end{align}
and called it {\em the individual Lindemann's index}.
This index is an extension of the Lindemann's index 
introduced in Subsec.~\ref{Subsec:Lindemann}.
The quantitative signature of surface melting is 
that the indices of surface atoms are significantly larger than 
those of inner atoms.

\begin{figure}[tbp]
\centering
\includegraphics[width=8cm]{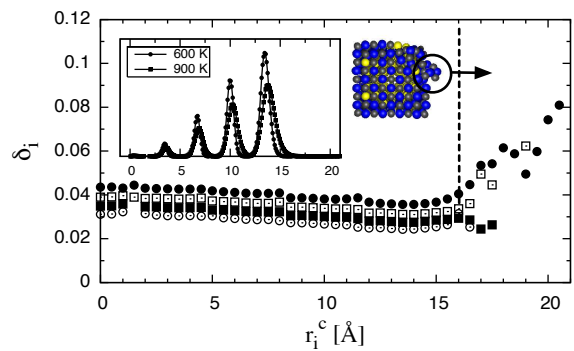}
  \caption{\footnotesize 
  Individual Lindemann's index $\delta_i$ in \KClBrxninemthree
  as a function of principal radial distance $\ri100$
    (white circles, solid squares, white squares and solid circles
    are the values at $600$, $700$, $800$ and $900$ K, 
    respectively).
  The frequency distribution of ions at $600$ and $900$ K 
  is imposed as a function of $\ri100$. }
  \label{atom-lindemann}
\end{figure}

Here, we show the MD result of {\SM} in \KClBrxninemthree as an example.
The mean square displacement between $i$-th and $j$-th ions,
$\ave{{r_{ij}}^2}-\ave{{r_{ij}}}^2$, 
which is averaged over a $10$ ps interval is computed.
At the same time the principal radial distance of $i$-th ion $\ri100$ was
also recorded.
Then one thousand samples of $i$-th ion's Lindemann index $\delta_i$ 
were obtained as a function of $\ri100$.
In Fig.~\ref{atom-lindemann}, we show $\delta_i$ 
as a function of $\ri100$ at $T = 600$, $700$, $800$ and $900$ K
averaged over the interval of $0.5\ \AA$.

It is evident that $\delta_i$ is $0.02$ - $0.04$ and 
almost constant over the entire cluster except 
the range $\ri100 > 16\ \AA$.
Increasing of $\delta_i$ in this range is so abrupt
that one might consider the surface is in a melting state.
However, it is not true due to the following reason:
almost all the surface ions of \KClBrxninemthree are 
at about $\ri100 = 14\ \AA$ so that 
there are no ions at the range $\ri100 > 16\ \AA$.
One of the insert figures in Fig.~\ref{atom-lindemann}
is the snapshot of the time evolution of $T = 900$ K at $3.48$ ns. 
At this moment, only five ions (two K$^+$ and three Br$^-$ ) 
were present on a surface of the cluster: ions whose positions $\ri100$ are
$19.3,\ 18.1,\ 16.8,\ 16.6$ and $20.8\ \AA$
and indeces  $\delta_i$ are 0.0782, 0.0538, 0.0391, 0.0729 and 0.0736,
respectively.
On the other hand, the other ions remain on their lattice sites.
Thus, the abrupt increasing of $\delta_i$ at the range $\ri100 > 16\ \AA$
is due to such a type of rare events.
One can also confirm this from
the other insert figure in Fig.~\ref{atom-lindemann} 
representing the distribution of ions as a function of $\ri100$.
The figure clearly shows that the relative frequency of ions
appearing at the range $\ri100 > 16\ \AA$ is significantly small.
Therefore, the large value of  $\delta_i$, which is attributed 
to exceptional atomic motion, has to be ignored 
when judging the presence of the surface melting.

Thus, the numerical results concerning the individual Lindemann's index
clearly show that the surface of \KClBrxninemthree  
is not in a melting state during {\SM}.
The absence of surface melting in alkali halide systems
is consistent with the result by Zykova-Timan {\it et.~al.}
\cite{Zykova-Timan:2005lr}.
It strongly suggests that the {\SPM} does not contribute to 
the intermixing of {\AH} clusters, unlike metal clusters.

\subsection{Characteristic motion of core ions}
\label{Subsec:core-atoms-behavior}

\begin{figure}[tbp]
\centering
\includegraphics[width=8cm]{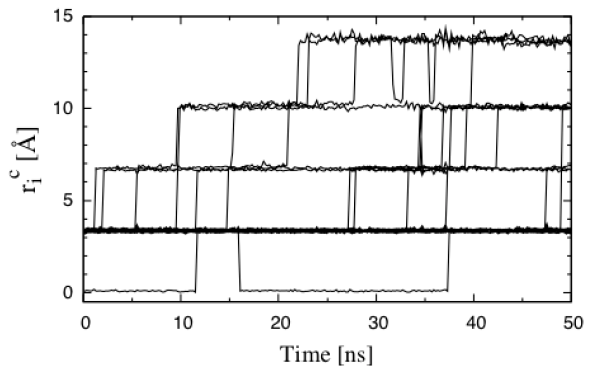}
  \caption{\footnotesize
 Typical time evolutions of radial distance of individual ions $\ri100$
 in \KClBrxninemthree corresponding to the run as depicted in
 Fig.~\ref{fig:KClBr-snapshot-all}~(c).
 }
 \label{fig:evolve-Ri}
\end{figure}

In this subsection, we illustrate the characteristic motion of core ions 
in \KClBrxninemthree during intermixing by observing the radial distance and 
by visualizing the projected positions of the ions.
We will come to the following fact: 
Both results demonstrate that diffusion of the ions 
is dominated by the {\VM}.

Let us summarize how the diffusion mechanism manifests 
in a characteristic motion of ions.
If either vacancy or interstitial mechanism is 
responsible for the intermixing of AH clusters, 
then core ions of the cluster have to diffuse individually and separately.
In contrast, the core ions tend to move simultaneously 
in case that the {\SPM} is responsible for {\SM}, 
because the cluster core is solid-like in the present simulation.
Thus, we can identify the  diffusion mechanism that governs the {\SM}  
by focusing on the motion of core ions in AH cluster.

\begin{figure}[tbp]
\centering
\includegraphics[width=8cm]{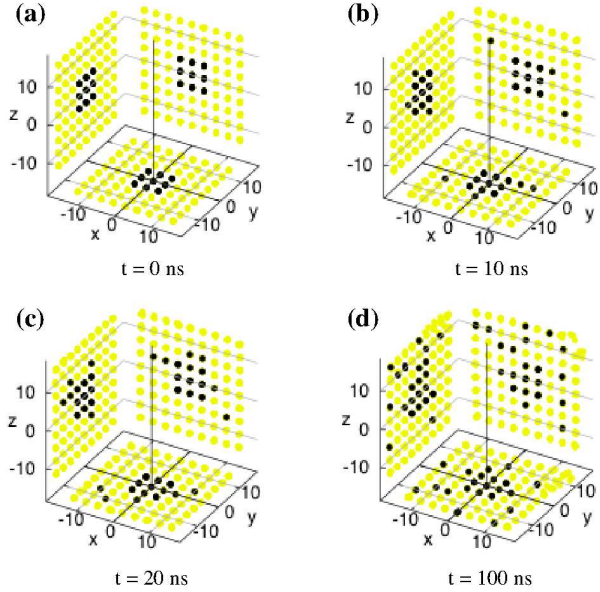}
  \caption{\footnotesize
  The snapshots of the ions of \KClBrxninemthree
  projected to $x$-$y$, $y$-$z$ and $z$-$x$ planes.  
  The time evolution corresponds to cases shown in
  Figs.~\ref{fig:KClBr-snapshot-all}~(c) and \ref{fig:evolve-Ri}.
  The black and yellow solid circles are the ions 
  on $3 \times 3 \times 3$ central cubic lattice sites at $t=0$
  and the other ions, respectively.
  }
 \label{fig:proj-snap-KClBr726}
\end{figure}

Next, we pick up the $27$ core ions which are initially located on 
$(3 \times 3 \times 3)$ lattice sites at the center of the cluster.
The initial configuration of $27$ core atoms is shown 
in Fig.~\ref{fig:proj-snap-KClBr726}~(a).
The radial distance between the $27$ core ions and the center, say $\ri100$, 
is computed during {\SM}.
Notice that $\ri100$ was previously introduced 
in Subsec.~\ref{Subsec:direct-observation}.
The time evolution of $\ri100$ for \KClBrxninemthree is
depicted in Fig.~\ref{fig:evolve-Ri}.

It is obvious that the plots corresponding to $\ri100=0\ \AA$ 
at $t=0$  keeps on the horizontal line during $t = 0 - 11$ and $17 - 36$ ns.
This means that the center ion yielding $\ri100=0\ \AA$ keeps its position
during these two periods.
In contrast, the plots corresponding to $\ri100=2.8\ \AA$ at $t=0$ 
tend to jump among the various horizontal lines. 
That indicates that $26$ ions which are initially belonging to core ions 
often move around from site to site. 
In other words, a single ion at the center stays 
at the same site for a long time, 
while the surrounding $26$ atoms often hop from one layer to another
individually and separately.
The present  behavior of the core 27 ions and the absence of surface melting
are incompatible with the description of {\SPM}.
Consequently, it strongly suggests that the diffusion inducing {\SM} 
is caused by the {\VM}.


We evaluate $\ri100=0\ \AA$ in the above, because in 3D model, 
it is difficult to visualize the motion of the hidden core atoms
directly in 3D space.
Here we visualize the diffusion process of interior ions of AH clusters 
by extracting the projected positions of ions onto three planes.
More specifically, we look into the projected coordinates of all atoms
on $x$-$y$, $y$-$z$ and $z$-$x$ planes.

In Fig.~\ref{fig:proj-snap-KClBr726},
we show snapshots of the projected positions of all the ions
which are obtained from MD data for \KClBrxninemthree during $t = 0$ - $100$ ns.
This corresponds to the time evolution shown in 
Figs.~\ref{fig:KClBr-snapshot-all}~(c) and \ref{fig:evolve-Ri}.

The position of the core ions are prepared to form
simple $(3 \times 3 \times 3)$ cubic  structure.
These core ions are initially located
in the vicinity of the center of the cluster. 
These initial core ions are depicted by black circles, whilst 
the other 699 ions surrounding 27 core ions are colored yellow.

The results show that several core ions (black solid circles)
diffuse and reach the cluster surface.
But most initial core ions remain at the center.
Moreover, there are no correlated motions of core ions in the snapshots.
These snapshots also support the previous observation indicating that 
the {\VM} is responsible for accomplishing {\SM} in AH clusters. 
Thus, {\SM} seems to originate from the combined effect
of the individual diffusion of ions realized by the {\VM}.

\subsection{Visualization of the vacancy motion}
\label{Subsec:vacancy-motion}

\begin{figure}[tbp]
\centering
\includegraphics[width=7cm]{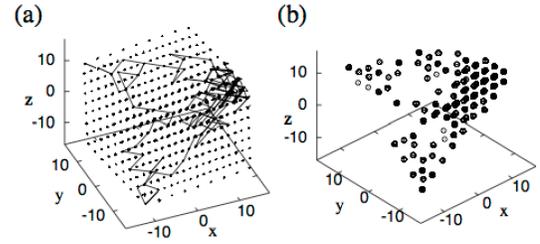}
 \caption{\footnotesize
 (a) Trajectories of all ions of \KClBrxninemthree 
 are plotted during $t = 38.1 - 39.4$ ns. 
 The line trajectories are due to hopping of ions.
 (b) All lattice sites occupied by a vacancy
 during $t = 38.1 - 39.4$ ns are displayed by circles.
}
 \label{fig:traj-all-atoms}
\end{figure}

A straightforward way to illustrate the crucial role of the {\VM} 
is to visualize the motion of a vacancy itself.
To do this, we put our focus on the events causing 
sudden changes of $\ave{r_i}$ in Fig.~\ref{fig:evolve-Ri}
which presents a visualization of atomic motion.
The trajectories of all the ions 
in \KClBrxninemthree 
are shown in Fig.~\ref{fig:traj-all-atoms}~(a)
during the period of $t=38.1 - 39.4$ ns.
A large number of dots and line segments form three zigzag lines.
One longer entangled line forms a single loop, and 
the other two relatively short lines are overlapped with the longer line
(the two shorter lines cannot be distinguished from the longer line).
The dots mean that ions stay for a long time on their lattice points.
The broken lines indicates the single hop of an ion from 
one occupied site to its second-neighbor unoccupied site,
which is nothing more than the so-called vacancy site.
Conversely, one can say
that each of the broken lines indicates a hopping of a vacancy 
in a direction opposite to the ion.
One can conclude from Fig.~\ref{fig:traj-all-atoms}~(a) that 
some ions hop from one site to another site 
so as to fill the vacancy created by a prior jump of  
an another ion. On the other hand,
almost all ions remain on the same lattice sites.
One can reconfirm this picture by comparing
Figs~\ref{fig:traj-all-atoms}~(a) and (b).
The circles in the latter figure indicate the lattice points occupied
by a vacancy in a particular period $t = 38.1 - 39.4$ ns.
The atomic trajectories depicted by the zigzag lines in the former
figure correspond to the vacancy sites depicted by circles
in the latter figure.
This figure makes manifest that
the atomic trajectory is equivalent to the vacancy trajectory.

It is hard to see the detail,
but there coexist three zigzag lines in Fig.~\ref{fig:traj-all-atoms},
corresponding to the trajectories of three vacancies 
in the cluster.
The starting points of the vacancies are located on
the edge of the initial configuration having a simple cubic structure.
As shown in Fig.~\ref{fig:init-conf}, 
these three vacancies are initially aligned along an edge.
In Fig.~\ref{fig:traj-all-atoms},
this vacancy trajectory forms a round trip loop ---
it starts from one of the surface vacancy sites,
wanders deep inside the cluster, and eventually reaches the surface again.

It takes a relatively short period of time for a vacancy to travel 
along the loop.
During the simulation time over $50$ ns, 
there appear many loops corresponding to the frequent
motion of vacancies everywhere inside of \KClBrxninemthree.
The formation of vacancy hopping paths inside the cluster,
is quite important to realize  intermixing in \KClBrxninemthree.
In this way, we are convinced that the {\SM} is dominated
by the {\em Vacancy Mechanism}.


\secti{Discussions}
\label{Sec:discussion}

The above numerical results gave evidence that
binary AH clusters can spontaneously intermix even below their melting points.
At a microscopic level, this intermixing is governed by 
migration of vacancies.
The vacancies mediated diffusion is by no means a process specific
to nanoclusters, but an ordinary process observed
also in bulk AH crystal. Although the transport process 
is the same, why is the mixing process in an AH cluster
much faster than in bulk AH? 
By analyzing the dominant factor controlling the diffusion rate 
in an AH cluster, we discuss the reason why the vacancy mechanism induces 
unexpectedly rapid diffusion in AH clusters.

\subsection{Diffusion rate in AH clusters}
\label{Subsec:diffusion-rate}

Firstly, to demonstrate how rapid {\SM} is,
we roughly evaluate the diffusion rate in {\AH} clusters
by using MD data \cite{yasuda1992saz,yasuda1992sos}.
Comparing the diffusion rate in cluster to that in bulk crystal
shows how fast the {\SM} is.
By comparing the diffusion rate in cluster to that in bulk crystal,
one might realize how fast the {\SM} is.
In general, it is difficult to estimate the diffusion rate in finite systems, 
where the surface diffusion coexists with bulk diffusion. 
Neglecting the spatial inhomogeneity  
due to the presence of the boundary, 
let us assume that diffusion rate is the same everywhere in the cluster.
This simplification enables us to estimate the diffusion rate
by $\tau$, which is the time 
spent by the guest atoms migrating over a distance, say $\Delta r_g'$.
Note, however, that such a crude estimation often
leads to  an underestimation.
The times required to diffuse over a distance 
$\Delta r_g' = 0.076 d\ \AA = 0.024\ \text{nm}$ 
at $700$ K in \KClBrxfivemthree, \KClBrxsevenmthree and \KClBrxninemthree
are evaluated as $\tau=$18.9, 42.1 and 668 ns, according to
the data in Fig.~\ref{fig:ave-inv-T-depend-KClBr}.
The resultant diffusion rates of the clusters are
$D = 3.1 \times 10^{-14},
1.4 \times 10^{-14}$ and 
$8.6 \times 10^{-16}\ \text{m}^2/\text{s}$, respectively.

Secondly, we estimate the rate of lattice diffusion 
in bulk AH crystal for comparison.
The diffusion rate by the {\VM} in solids is usually represented by
\begin{align}
D = D_0 \exp \left[ - {(E_f + E_m)}/{k_B T} \right],
 \label{vacancy-diff-const-eq}
\end{align}
where $D_0$ is a constant factor, and
$E_f$ and $E_m$ are the vacancy formation and migration energy, respectively.
The sum of $E_f$ and $E_m$ is the activation energy of vacancy diffusion
$E_{act}$.
For instance, the energies in bulk KCl and NaCl crystals are
approximately $E_f^S \simeq 2$ eV and $E_m \simeq 0.9$ eV, 
respectively \cite{Kittel2005ed8th}.
Since $E_f^S$ is Schottky vacancy formation energy 
for an anion-cation pair,
the formation energy of a single vacancy on an anion site 
is roughly estimated to be a half of $E_f^S$, i.e., $E_f \simeq 1$ eV.
For {\it binary} AH crystals containing two different types of
anions such as KCl-KBr, only a limited number of experimental 
studies have been done to exploring the atomic diffusion. 
In particular, based on experimental results, Pehkonen and his coworkers 
estimated the activation energy in KCl-KBr powder mixtures 
as $E_{act} = 2.1 \pm 0.1$ eV
\cite{Pehkonen1972KClBrActE,Pehkonen1973DiffKCl-Br}.
According to their results, the activation energy of bulk AH crystals
is approximately $2$ eV.
The constant factor is also roughly estimated as 
$D_0 = 1 \times 10^{-4}\ \text{m}^2/\text{sec}$
by experiments \cite{LaurancePhysRev57NaClDiff,MizunoPhysRev1226KClDiff}.
Then, by substituting these values into Eq.~(\ref{vacancy-diff-const-eq})
we obtain the diffusion rate in bulk KCl-KBr crystals at $700$ K;
$D_{bulk} \simeq 4.0 \times 10^{-19}\ \text{m}^2/\text{s}$.
In addition, we also obtain the diffusion time $\tau$ as follows;
$\tau = 1.5 \times 10^{6}\ \text{ns}$ for $\Delta r_g' = 0.076 d\ \AA$.

The value $\tau$ in the clusters $\tau = 18.9$-$668$ ns
is at least about four orders of magnitude less than $\tau$ in bulk solids.
The diffusion rate $D$ in the clusters is, at least, about 
two orders of magnitude larger than the one in bulk.
The activation energy of the diffusion $0.98$ - $1.4$ eV
estimated in Sec.~\ref{Subsec:T-depend} is significantly lower than
that of the bulk diffusion.
Taking into account these results, it is reasonable to conclude 
that the {\SM} in the MD simulations is a significantly more rapid diffusion.


\subsection{Principal origin of enhanced intermixing}
\label{Subsec:mechanism}

Up to now, 
the reason why the diffusion rate in AH cluster is significantly
more rapid than in  AH bulk has not been clarified.
At first sight, there seems to be no qualitative 
difference between the diffusion in bulk and that in a cluster.
In this subsection, we discuss the factor enhancing
the diffusion in AH clusters from the viewpoint of 
the size-dependence of {\em vacancy formation energy} $E_f$.
In particular, we will demonstrate a decreasing trend of the formation 
energy with reducing size of AH cluster. 
This provides evidence
that reduction of the cluster size makes it energetically
easier to generate vacancies causing SM.

The diffusion rate controlling the {\SM} is determined 
by the vacancy formation energy and the migration energy
as shown in Eq.~(\ref{vacancy-diff-const-eq}).
It is plausible to expect that the values of these energies
depend upon the cluster size like other 
physical properties exhibiting the size effect.
In fact, there are several theoretical studies which point out 
that the vacancy formation energy $E_f$ in metal clusters depends on the size
in the metal clusters \cite{PhysRevB.50.2775,qi2003sdv,muller2007ctv}.
But it has not been well studied
how the size effect manifests itself in AH clusters.
In order to compute the value $E_m$, 
it is necessary to enumerate the reaction paths along which the vacancy hops.
Since there also exists a large variety of the saddle points
which may contribute to the numerical estimation for $E_m$,
it is practically hard to calculate $E_m$.
Rather than pursuing the role of $E_m$, in this paper
we concentrate on the numerical estimation of $E_f$.
Here we would like to
extract an insight about the effect of size
upon the enhancement of diffusion 
by focusing on the size-dependent feature of $E_f$.

According to the definition of the vacancy formation energy,
$E_f$ is the energy which is spent for the process 
such that a single constituent atom is removed from the target system 
to the surface.
Let us apply that definition to calculate $E_f$ of
clusters which have the same number of atoms but different cluster shapes.
The value $E_f$ would depend on the shapes
even if the number of the constituent atoms is the same,
because the energy needed to put a single atom on the surface
depends upon the detailed structure of the surface.
To avoid the difficulty of accounting for the complicated 
shape-dependence of the cluster,
we evaluated $E_f$ of KCl-KBr clusters for a series of the clusters
forming a simple cubic structure represented by $(n \times n \times n)$,
locating the vacancy at the center of a cluster.
In addition, we applied the following convenient relation 
to evaluate $E_f$ in a cluster composed of $N$ atoms:  
\begin{align}
E_f(N) = E(N-1,v) - \frac{N-1}{N} E(N,0),
\label{eq:VFE}
\end{align}
where $E(N-1,v)$ and $E(N,0)$ are the energy of a cluster 
including $(N-1)$ and $N$ atoms with and without a vacancy,
respectively \cite{PhysRevB.50.2775,muller2007ctv}.

There are some issues about the applicability 
of the above formula to a cluster. 
Strictly speaking, the formation energy of a vacancy, $E_f$, should be 
defined as the difference between the energy of a cluster with a vacancy 
on the most stable surface position and the energy with the vacancy
at a position inside the cluster.
However, the latter energy depends on the vacancy's position inside the cluster.
Thus, the effective vacancy formation energy should be estimated 
by averaging $E_f$s over all the lattie points inside the cluster.
The validity of this estimation could be verified by comparing 
it with another vacancy formation energy which is  evaluated 
by MD simulations.
We will discuss such quantitative verification of $E_f$ 
in clusters in a forthcoming paper.
In this paper we only present a rough treatment based on Eq.~(\ref{eq:VFE}).

\begin{figure}[tbp]
\centering
\includegraphics[width=8cm]{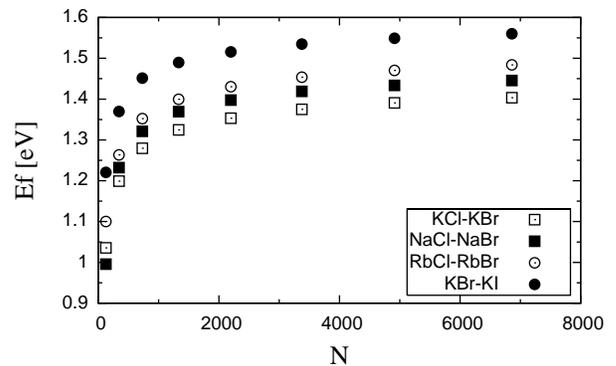}
 \caption{\footnotesize
  Size dependence of the vacancy formation energy  $E_f$ in 
 KCl-KBr, NaCl-NaBr, RbCl-RbBr and KBr-KI clusters, where $N$ is the
 number of constituent ions.
 }
 \label{fig:VFE}
\end{figure}

In Fig.~\ref{fig:VFE} 
we show the size-dependence of $E_f$ in various kind of AH clusters;
KCl-KBr, NaCl-NaBr, RbCl-RbBr and KBr-KI clusters.
It is obvious that $E_f$ for all kind of AH clusters decreases
remarkably with decreasing size \cite{Niiyama2011SASMLett}.
Thus the diffusion by vacancies is further
enhanced as the size decreases.
This is the origin of the size-dependence of activation
energy and it is an underlying reason why the Vacancy 
Mechanism effectively accelerates the SM in a smaller cluster.
It is expected that this size-dependence results from
the Coulomb long range force interacting between ions in {\AH} clusters.
More details for the origin of the size-dependence of vacancy 
formation energy will be discussed in a forthcoming paper.


\secti{Conclusion}

Rapid diffusion in AH nano clusters,
so-called {\em Spontaneous Mixing},
and its atomistic mechanism were investigated 
by Molecular Dynamics simulations employing a three dimensional
model for ternary alkali halides: KCl-KBr, NaCl-NaBr, RbCl-RbBr and KBr-KI. 

To quantify the atomistic mixing process 
we introduced an indicator, $\inv$, which is a good indicator of
the degree of intermixing in a cubic shape cluster.
Applying this indicator to the time evolutions of 
\KClBrxfivemthree, \KClBrxsevenmthree and \KClBrxninemthree
and observing their snapshots, 
we confirmed that the SM is surely reproduced in our model clusters.
On the other hand, by computing Lindemann's
index, which measures the relative fluctuation
of the distance between any two atoms,
we proved that the intermixing process proceeds in the solid phase.
Similar results were also
obtained for AH clusters with different combination of species
such as \NaClBrxfivemthree, \RbClBrxfivemthree and \KBrIxfivemthree.

The analysis of MD simulations at $600$-$900$ K indicated the 
Arrhenius type temperature-dependence of the time required
for the intermixing.
By fitting with Arrhenius law, we also estimated 
the activation energy: $E_{act} = 0.98, 1.1$ and $1.4$ eV
for \KClBrxfivemthree, \KClBrxsevenmthree and \KClBrxninemthree, respectively.
It was also confirmed that
the activation energy as well as 
the intermixing time depend on the size of the clusters.

To clarify the atomistic mechanism of the diffusion,
we scrutinized surface melting 
and the dynamics of ions inside the AH clusters during the intermixing.
A detailed evaluation of individual Lindemann's index $\delta_i$
reveals a very important fact: in contrast to the metal nano clusters 
the surface melting, which is the necessary condition 
for Surface Peeling Mechanism, was not observed at all in \KClBrxninemthree.
At the same time, we observed that the time evolutions of radial 
distance $\ri100$ of inner ions together with the snapshots of 
the position of all ions projected onto three planes also 
do not show any indication of collective motions peculiar to the {\SPM}.
All the examples studied in our simulations 
indicate individual diffusive motions 
rather than collective motions dominate the mixing process.
By visualizing the trajectories of ions 
we found the {\VM} is responsible for the intermixing in AH clusters.

The rough estimation of the diffusion rate $D$ of the intermixing
in \KClBrxfivemthree, \KClBrxsevenmthree and \KClBrxninemthree at $700$ K
gave $D= 3.1 \times 10^{-14}$, $1.4 \times 10^{-14}$, 
and $8.6 \times 10^{-16}$ m$^2/$s.
These rates are much larger than the diffusion constant
of bulk alkali halide crystals $\sim 4 \times 10^{-19}$ m$^2/$s. 
These facts support the conclusion 
that the intermixing observed in our MDs is 
rapid diffusion peculiar to nano particles.

Finally, evaluation of vacancy formation energy $E_f$ for 
AH clusters with different species and sizes showed that 
$E_f$ definitely decreases as the size of the clusters becomes small, 
which is consistent with the size dependence of activation energy.
This decrease of vacancy formation energy with size is the principal
origin of the rapidness of {\SM} in AH clusters.

\begin{acknowledgments}
The authors thank M. Ohnishi and  M. Watanabe  for their preliminary 
calculation in the early stage of the research. 
This work is partly supported by JSPS KAKENHI(Grant No.23540459 and 08J08311).
\end{acknowledgments}

\bibliography{/Users/niyama/Documents/Articles}

\end{document}